\documentclass[conference]{IEEEtran}
\usepackage{algorithm}
\usepackage[noend]{algpseudocode}
\usepackage{fancyhdr}

\usepackage{epsfig,latexsym}
\usepackage{float}
\usepackage{indentfirst}
\usepackage{amsmath}
\usepackage{amssymb}
\usepackage{times}
\usepackage{subfigure}
\usepackage{pifont}
\usepackage{psfrag}
\usepackage{cite}
\usepackage{url}
\usepackage{lastpage}
\usepackage{bm}
\usepackage{color}
\usepackage{fancyhdr}
\usepackage[center]{caption2}
\usepackage{balance}

\newtheorem{Proposition}{Proposition}

\ifCLASSINFOpdf  
\else
\fi

\begin{document}
%
\title{Secure Energy Efficiency Optimization for MISO Cognitive Radio Network with Energy Harvesting}

\author{\IEEEauthorblockN{Miao Zhang，, Kanapathippillai Cumanan and Alister Burr }
\IEEEauthorblockA{Department of Electronic Engineering, University of York, York, YO10 5DD, United Kingdom\\
Email: \{mz1022, kanapathippillai.cumanan, alister.burr\}@york.ac.uk}}

\maketitle

\begin{abstract}
	This paper investigates a secure energy efficiency (SEE) optimization problem in a multiple-input single-output (MISO) underlay cognitive radio (CR) network. In particular, a multi-antenna secondary transmitter (SU-Tx) simultaneously sends secured information and energy to a secondary receiver (SU-Rx) and an energy receiver (ER), respectively, in the presence of a primary receiver (PU-Rx). It is assumed that the SU-Rx, ER and PU-Rx are each equipped with a single antenna. In addition, the SU-Tx should satisfy constraints on maximum interference leakage to the PU-Rx and minimum harvested energy at the ER. In this CR network, we consider the transmit covariance matrix design with the assumption of perfect channel state information (CSI) at the SU-Tx. In addition, it is assumed that the ER is a potential passive eavesdropper due to broadcast nature of wireless transmission. On the other hand, we consider the worst-case scenario that ER's energy harvesting requirement is only satisfied when it performs only energy harvesting without intercepting or eavesdropping information intended for the SU-Rx. We formulate this transmit covariance matrix design as a SEE maximization problem which is a non-convex problem due the non-linear fractional objective function. To realize the solution for this non-convex problem, we utilize the non-linear fractional programming and difference of concave (DC) functions approaches to reformulate into a tractable form. Based on these techniques and the Dinkelbach's method, we propose iterative algorithms to determine the solution for the original SEE maximization problem. Numerical simulation results are provided to demonstrate the performance of the proposed transmit covariance matrix design and convergence of the proposed algorithms.
\end{abstract}
\IEEEpeerreviewmaketitle
\begin{IEEEkeywords}
	Secure energy efficiency (SEE), energy harvesting, MISO, convex optimization. 
\end{IEEEkeywords}

%
\IEEEpeerreviewmaketitle

\section{Introduction}
Without doubt, information security is one of the most critical issues of wireless communications due to the open nature of transmission over the wireless medium. Traditionally, information security techniques are implemented at the application layer based on cryptographic techniques which mainly rely on the computational complexity of difficult mathematical problems \cite{liang}. On the other hand, the broadcast nature of wireless communications introduces different challenges in terms of key exchange and distribution \cite{chu1,cumanan2014secrecy,cumanan2016secrecy}. Information theoretic studies have proven that if the signal to noise ratio (SNR) of the legitimate channel is larger than that of the eavesdropper's channel, secure communication can be guaranteed \cite{Shannon}, which is known as physical layer security in the literature. This approach was first theoretically proposed by Shannon \cite{Shannon} and then the secrecy capacity of wiretap and related channels were developed by Wyner \cite{Wyner} and Csiszar \cite{csiszar}. Physical layer security exploits physical layer characteristics of wireless channels including randomness to achieve secure communication between legitimate parties in the presence of eavesdroppers \cite{li2017optimal,cumanan2010multiuser}. In contrast to conventional security techniques, physical layer security has lower computational complexity for practical implementation \cite{cumanan2017physical,cumanan2017secure}.\\
Achieving higher data rate, energy efficiency and information security are the essential requirements of future wireless communications, including fifth generation (5G) wireless networks. However, with the exponential growth of the number of wireless devices with high data rate and security requirements, energy consumption has become one of the critical issues in terms of both environmental and economic aspects \cite{hu2016optimal}. In addition, wireless communications consume two percent of the entire world energy \cite{zappone2014energy}, and this percentage will grow rapidly with the increasing number of wireless devices and the development of new communication technologies. This growth in energy consumption will result in more carbon emission and electromagnetic pollution to the environment. In addition, due to the limited battery life of mobile devices and slow development of energy storage technologies, energy efficient communications have recently become a promising approach to address these issues. \\
Most work on physical layer security in the literature is either secrecy rate maximization with a total transmit power constraint \cite{zhang2016secrecy,yuan2014joint,chu2016secrecy,chu2014secrecy} or power minimization to meet the secrecy rate requirements \cite{li2011cooperative,chu2016simultaneous,chu2015robust}. However, the solutions for the above mentioned optimization problems might not be able to achieve the optimal SEE, as the objective functions of these problems are optimized while satisfying the constraints. Therefore, we consider the SEE based resource allocation problem in this paper to measure efficient utilization of transmit power in a secure communication system. The SEE is defined as the ratio between the achievable secrecy rate and the total transmit power consumption.
\\
Wireless energy harvesting (EH) is a newly emerging technique to harvest energy from the information carrying radio frequency signals radiated from transmitters \cite{varshney,grover2010shannon}. Conventional EH methods usually collect energy from the external natural sources, like wind, solar, etc \cite{hossain2014evolution,raghunathan2006emerging}. However, these external energy resources are not constantly stable and are difficult to apply to mobile devices, for example, the size of harvesting devices and the geographical limitations. In comparison to other renewable energy sources such as solar and wind, wireless EH is easier to implement and design for mobile devices \cite{zhou2013wireless}.\\
Motivated by the aforementioned aspects, we investigate the SEE maximization problem for an underlay MISO CR network with EH requirement. In particular, a multi-antenna SU-Tx simultaneously sends secured information and energy to a SU-Tx and ER, respectively, in the presence of a PU-Rx as shown in Fig. 1. We consider transmit covariance matrix design to maximize the achievable SEE with secrecy rate on the SU-Rx, interference leakage \cite{cumanan2009sinr} and EH requirement. On the other hand, the ER is considered to be a potential passive eavesdropper due to broadcast nature of wireless transmission. With the perfect channel state information (CSI) assumption, we formulate the transmit covariance matrix design problem to maximize SEE under these constraints. The original SEE maximization problem is not convex due to its non-linear fractional objective function and it introduces some challenges in realizing the solution. To circumvent this issue, we reformulate this problem into a tractable form by exploiting non-linear fractional programming \cite{dinkelbach1967nonlinear} and difference of concave (DC) functions programming \cite{dinh2014recent}. Though the reformulated problem is still non-convex, we show that the optimal solution can be obtained by iteratively solving the problem with the help of \emph{non-linear fractional} programming and DC programming. 
\\
The remainder of this paper is organized as follows. The system model is presented in Section II, and the SEE maximization problem with the perfect CSI assumption is formulated and iterative algorithms are proposed to solve it in section III. Section IV provides simulation results to validate the performance of the proposed algorithms and finally Section V concludes this paper. 
\subsection{Notations}
We use upper and lower case boldface letters for matrices and vectors, respectively. $(\cdot)^{-1}$, $(\cdot)^T$ and $(\cdot)^H$ stand for the inverse, transpose and conjugate transpose operations, respectively. $\mathbf{A}\succeq\mathbf{0}$ means that $\mathbf{A}$ is a positive semidefinite matrix. $\textrm{rank}(\mathbf{A})$ denotes the rank of a matrix, and $\textrm{tr}(\mathbf{A})$ represents the trace of matrix $\mathbf{A}$. The circularly symmetric complex Gaussian (CSCG) distribution is represented by $\mathcal{CN}(\mu,\sigma^2)$ with mean $\mu$ and variance $\sigma^2$. $\mathbb{H}^{N}$ denotes the set of all $N \times N$ Hermitian matrices.\\
\section{System Model}
\begin{center}
	\begin{figure}[ht!]
		\includegraphics[width=\linewidth]{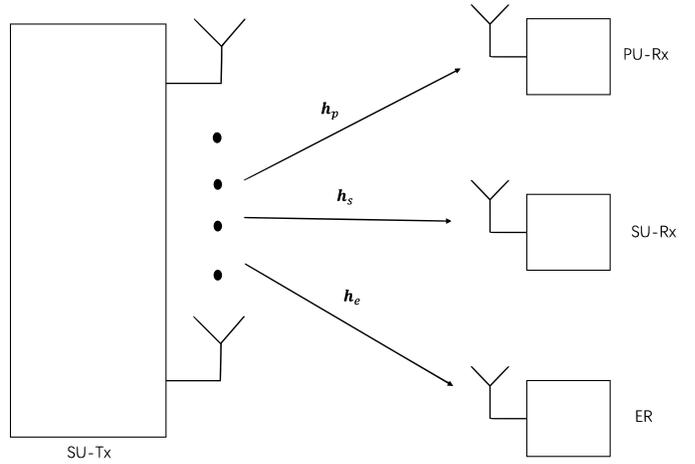}
		\caption{An underlay CR network with a multi-antenna SU-Tx and PU-Rx, SU-Rx and ER are equipped with single antenna. }	
		\label{fig:SRP}
	\end{figure}
\end{center}
 We consider a MISO CR network with four terminals: one SU-Tx, one SU-Rx, one PU-Rx and one ER. The SU-Tx intends to send confidential message to the SU-Rx while the interference leakage to the PU-Rx should not exceed a predefined threshold. The ER harvests energy from the SU-Tx through wireless power transfer. However, a potential issue might arise that the ER might turn out to be a potential eavesdropper and attempts to intercept the message sent to the SU-Rx. Therefore, it is assumed that the ER is a passive eavesdropper in this CR network. We focus on the worst-case scenario that the SU-Tx guarantees the EH requirement only when the ER does not attempt to decode the message \cite{leng2014power}. The SU-Tx is equipped with $N_{t}$ antennas, while the ER, the SU-Rx and the PU-Rx each have only a single antenna. The channel coefficients between the SU-Tx and the PU-Rx, the SU-Rx and the ER are denoted by $\mathbf{h}_{p}\in\mathcal{C}^{N_{T}\times1}$, $\mathbf{h}_{s}\in\mathcal{C}^{N_{T}\times1}$ and $\mathbf{h}_{e}\in\mathcal{C}^{N_{T}\times1}$, respectively. Thus, the received signal at SU-Rx and ER can be written as
\begin{align}
y_{s}=\mathbf{h}_{s}^{H}\mathbf{x}+n_{s},\\
{y}_{e}=\mathbf{h}_{e}^{H}\mathbf{x}+{n}_{e},
\end{align}
where $\mathbf{x}\in\mathcal{C}^{N_{T}\times1}$ denotes the transmitted signal from the SU-Tx, whose
transmit covariance matrix is defined as $\mathbf{Q}_{s} (\succeq 0)=E(\mathbf{x}\mathbf{x}^{H})\in\mathcal{C}^{N_{T}\times N_{T}}$. $n_{s}\sim\mathcal{CN}(0,1)$ and ${n}_{e}\sim\mathcal{CN}(0,1)$ denote the additive white Gaussian noise (AWGN) at the SU-Rx and the ER, respectively. For guaranteeing communication security, we consider the worst-case scenario that the ER can only harvest energy when it does not attempt to eavesdrop the SU-Rx message. Denote $R_{s}$ as the achievable secrecy rate of SU-Rx:
\begin{align}
R_{s}=\log_{2}(1+\mathbf{h}_{s}^{H}\mathbf{Q}_{s}\mathbf{h}_{s})-\log_{2}(1+\mathbf{h}_{e}^{H}\mathbf{Q}_{s}\mathbf{h}_{e}).
\end{align}
The total transmit power consumption at SU-Tx is given by:
\begin{align}
P_{t}=\frac{\textrm{tr}(\mathbf{Q}_{s})+P_{c}}{\xi},
\end{align}
where $P_{c}$ is the circuit power consumption of the transmitter and $\xi\in(0,1]$ is the power amplifier efficiency, which is assumed to be one  ($\xi=1$) without loss of generality in this paper. The SEE is defined as the ratio between the achievable secrecy rate and the total transmit power consumption, which can be written as
\begin{align}
\eta=\frac{R_{s}}{P_{t}}=\frac{\log_{2}(1+\mathbf{h}_{s}^{H}\mathbf{Q}_{s}\mathbf{h}_{s})-\log_{2}(1+\mathbf{h}_{e}^{H}\mathbf{Q}_{s}\mathbf{h}_{e})}{\textrm{tr}(\mathbf{Q}_{s})+P_{c}}.
\end{align}
 The harvested energy at ER can be defined as
\begin{align}
E_{eh}=\eta_{eh}(\mathbf{h}_{e}^{H}\mathbf{Q}_{s}\mathbf{h}_{e}+1),
\end{align}
where $\eta_{eh}\in(0,1]$ is the energy conversion ratio at the ER. \\
\section{Problem Fomulation}
 In this section, we solve a SEE maximization problem with the constraints on the minimum harvested energy at ER and the maximum interference leakage at the PU-Rx. This SEE maximization problem can be formulated as
 \begin{subequations}\label{eq:Sec_rate_max_ori}
 	\begin{align}
 	\max_{\mathbf{Q}_{s}} &~\eta=\frac{\log_{2}(1+\mathbf{h}_{s}^{H}\mathbf{Q}_{s}\mathbf{h}_{s})-\log_{2}(1+\mathbf{h}_{e}^{H}\mathbf{Q}_{s}\mathbf{h}_{e})}{\textrm{tr}(\mathbf{Q}_{s})+P_{c}}, \label{eq:Sec_rate_max_obj} \\
 	s.t. &~R_{s}=\log_{2}(1+\mathbf{h}_{s}^{H}\mathbf{Q}_{s}\mathbf{h}_{s})-\log_{2}(1+\mathbf{h}_{e}^{H}\mathbf{Q}_{s}\mathbf{h}_{e})\geq R_{d}\\
 	&~  E_{eh}=\eta_{eh}(\mathbf{h}_{e}^{H}\mathbf{Q}_{s}\mathbf{h}_{e}+1) \geq E_{s},  \label{eq:Sec_rate_max_EH_contraints}\\
 	&~ \mathbf{h}_{p}^{H}\mathbf{Q}_{s}\mathbf{h}_{p}\leq P_{f},  \textrm{tr}(\mathbf{Q}_{s}) \leq P_{\textrm{tx}}, \mathbf{Q}_{s}\succeq 0.  \label{eq:Sec_rate_max_power_constraints}
 	\end{align}
 \end{subequations}
 The physical meaning of the constraint in (7c) is that the transmitter should satisfy the minimum harvest energy requirement at the ER if it is only interested in EH and not in eavesdropping the SU-Rx signal. This problem is not a convex problem due to the fractional objective function, and we convert this problem into a convex one through non-linear fractional and DC programming in the following subsections.
 \subsection{Non-linear fractional programming}
 The objective function in (7a) is a fractional programming problem with non-linear as well as linear terms in the numerator and denominator, therefore the problem in (7) is known as a non-linear fractional problem in literature \cite{dinkelbach1967nonlinear}. The original problem can be converted into a parametric programming problem \cite{dinkelbach1967nonlinear}.  Denote 
 \begin{align}
 \lambda^{*}=\frac{R_{s}^{*}}{P_{t}^{*}},
 \end{align}
 where $R_{s}^{*}$ and $P_{t}^{*}$ are the optimal secrecy rate and power consumption of problem (7), respectively. The maximum SEE 
 \begin{align}
 \lambda^{*}=\frac{R_{s}^{*}}{P_{t}^{*}}=\max_{\mathbf{Q}_{s}}\frac{R_{s}}{P_{t}}
 \end{align}
 can be achieved only when $\lambda^{*}$, ${R_{s}^{*}}$ and $P_{t}^{*}$ satisfy the following condition \cite{dinkelbach1967nonlinear}
 \begin{align}
 \max_{\mathbf{Q}_{s}}~[R_{s}-\lambda^{*}P_{t}]=R_{s}^{*}-\lambda^{*}P_{t}^{*}=0,
 \end{align}
 for $R_{s}\geq 0$ and $P_{t}>0$. The parametric programming problem with parameter $\lambda$ is defined as
 \begin{align}
\max_{\mathbf{Q}_{s}}~[R_{s}-\lambda P_{t}]=&\max_{\mathbf{Q}_{s}}\{\log_{2}(1+\mathbf{h}_{s}^{H}\mathbf{Q}_{s}\mathbf{h}_{s})-\nonumber\\
&~\log_{2}(1+\mathbf{h}_{e}^{H}\mathbf{Q}_{s}\mathbf{h}_{e})-\lambda[ \textrm{tr}(\mathbf{Q}_{s})+P_{c}]\}\nonumber\\
s.t. &~\textrm{(7b)-(7d)}. 
\end{align}
 It can be seen that the original problem (7) is transformed into a parameterized polynomial subtractive form. As a result, the original problem is reformulated to find $\lambda^{*}$ and $\mathbf{Q}_{s}^{*}$ to satisfy the condition in (10). By utilizing Dinkelbach's method \cite{dinkelbach1967nonlinear} with an initial value $\lambda_{0}$ of $\lambda$, the optimal solutions of (11) can be obtained iteratively by solving
 \begin{align}
 \max_{\mathbf{Q}_{s}}&~[R_{s}-\lambda_{i} P_{t}]\nonumber\\
 s.t.&~\textrm{(7b)-(7d)}. 
 \end{align}
  with a given $\lambda_{i}$ at the $i$th iteration, where $i$ is the iteration index. $\lambda_{i}$ can be considered as the SEE obtained at the previous iteration. At each iteration, $\lambda_{i}$ should be updated as 
 \begin{align}
 \lambda_{i+1}=\frac{R_{s}^{i}}{P_{t}^{i}},
 \end{align}
 where $R_{s}^{i}$ and $P_{t}^{i}$ denote the solution of (12) for the given $\lambda_{i}$. This iterative process will be terminated when the condition in (10) is satisfied. However, in practice the iterative process will be repeated until the following inequality is satisfied:
 \begin{align}
 \Delta F=|R_{s}^{i}-\lambda_{i}P_{t}^{i}|\leq \varepsilon,
 \end{align}
 with a small convergence tolerance $\varepsilon>0$. The proposed algorithm of non-linear fractional programming is summized in Algorithm 1.
 \begin{algorithm}
 	\caption{Non-linear fractional programming}\label{alg:euclid}
 	\begin{algorithmic}[1]
 		\State Initial $i=0$ and choose an initial value $\lambda_{0}$;
 		\Repeat 
 		\State For the given $\lambda_{i}$, find the optimal $R^{i}_{s}$ and $P_{t}^{i}$ of (12) (DC programming);
 		\State Update  $\lambda_{i+1}=\frac{R_{s}^{i}}{P_{t}^{i}}$ to obtain $\lambda_{i+1}$ 
 		\State $i=i+1$;
 		\Until (14) satisfied;
 		\State Return $\lambda^{*}=\lambda_{i}, P^{*}_{t}=P_{t}^{i-1}, R_{s}^{*}=R_{s}^{i-1}$.
 	\end{algorithmic}
 \end{algorithm}
 \subsection{DC programming}
 DC programming is an optimization approach to solve non-convex problems. In particular, this technique can be applied for an optimization problem with an objective function, which is a difference of two concave functions. Since, the objective function in (12) falls into this category, DC programming can be utilized to solve this problem.\\
 The fundamental idea of DC programming \cite{dinh2014recent} is to locally linearize the non-concave functions at a feasible point $\mathbf{Q}_{s}^{k}$ and then iteratively solve the linearized problem. We define the following function to approximate the second term of the objective function in (12)
 \begin{align}
 f(\mathbf{Q}_{s},\mathbf{Q}_{s}^{k})=\log_{2}(1+\mathbf{h}_{e}^{H}\mathbf{Q}_{s}^{k}\mathbf{h}_{e})+\frac{\mathbf{h}_{e}^{H}(\mathbf{Q}_{s}-\mathbf{Q}_{s}^{k})\mathbf{h}_{e}}{(1+\mathbf{h}_{e}^{H}\mathbf{Q}_{s}^{k}\mathbf{h}_{e})\ln2}.
 \end{align}
 Based on this approximation, the problem (12) can be converted into the following equivalent problem:
 \begin{align}
 \max_{\mathbf{Q}_{s}}~&\{\log_{2}(1+\mathbf{h}_{s}^{H}\mathbf{Q}_{s}\mathbf{h}_{s})-f(\mathbf{Q}_{s},\mathbf{Q}_{s}^{k})-\lambda_{i}[ \textrm{tr}(\mathbf{Q}_{s})+P_{c}]\}\nonumber\\
 s.t. &~1+\mathbf{h}_{s}^{H}\mathbf{Q}_{s}\mathbf{h}_{s}\geq 2^{R_{d}}(1+\mathbf{h}_{e}^{H}\mathbf{Q}_{s}\mathbf{h}_{e}),\nonumber\\
 &~  \eta_{eh}(\mathbf{h}_{e}^{H}\mathbf{Q}_{s}\mathbf{h}_{e}+1) \geq E_{s},\nonumber \\
 &~ \mathbf{h}_{p}^{H}\mathbf{Q}_{s}\mathbf{h}_{p}\leq P_{f}, \textrm{tr}(\mathbf{Q}_{s}) \leq P_{\textrm{tx}}, \mathbf{Q}_{s}\succeq 0. 
 \end{align}
 This approximated problem is convex in terms of $(\mathbf{Q}_{s})$ and hence $\mathbf{Q}_{s}^{*}$ can be obtained through iteratively solving problem (16) and iteratively updating $\mathbf{Q}_{s}^{k}$. The algorithm based on DC programming is provided in Algorithm 2.
 \begin{algorithm}
 	\caption{DC Programming}\label{alg:DC}
 	\begin{algorithmic}[1]
 		\State Initial $k=0$, choose an initial value $\mathbf{Q}_{s}^{k}=0$ and $\eta^{i,k}=0$;
 		\Repeat 
 		\State Solve the problem (16) with $\lambda=\lambda_{i}$ from Algorithm 1  and obtain $\mathbf{Q}_{s}^{k+1}$;
 		\State Compute  $\eta^{i,k+1}=\log_{2}(1+\mathbf{h}_{s}^{H}\mathbf{Q}_{s}^{k+1}\mathbf{h}_{s})-\log_{2}(1+\mathbf{h}_{e}^{H}\mathbf{Q}_{s}^{k+1}\mathbf{h}_{e})-\lambda_{i}[\textrm{tr}(\mathbf{Q}_{s}^{k+1})+P_{c}]$ ;
 		\State $\Delta\eta=\eta^{i,k+1}-\eta^{i,k}$;
 		\State Update $k=k+1$;
 		\Until $|\Delta\eta|\leq \zeta$;
 		\State Return $R_{s}^{i}=\log_{2}(1+\mathbf{h}_{s}^{H}\mathbf{Q}_{s}^{k}\mathbf{h}_{s})-\log_{2}(1+\mathbf{h}_{e}^{H}\mathbf{Q}_{s}^{k}\mathbf{h}_{e})$ and $P_{t}^{i}=\textrm{tr}(\mathbf{Q}_{s}^{k})+P_{c}$ to Algorithm 1 for updating $\lambda_{i+1}$.
 	\end{algorithmic}
 \end{algorithm}
\begin{Proposition}\label{proposition:rank_proof}
	Provided that the problem (11) is feasible, the optimal solution will be always rank-one.
\end{Proposition}
\begin{IEEEproof}
	Please refer to Appendix.
\end{IEEEproof}
\section{Simulation Results}
In this section, we provide numerical simulation results to validate the performance of the proposed schemes. The SU-Tx is equipped with three ($N_{t} = 3$) antennas, while the PU-Rx, SU-Rx and ER each use a single antenna. All the channel coefficients are generated by CSCG with zero mean and unit variance. The maximum interference leakage to the PU-Rx is assumed to be 0 dB. In addition, the energy conversion ratio is assumed to be 0.5. The convergence tolerances $\varepsilon$ and $\zeta$ are set to be $10^{-3}$.\\
First, we evaluate the convergence of the proposed algorithms in Fig. 2 for the target secrecy rate $R_{d} = 0.5$ bps/Hz, the power consumption for transmission $P_{\textrm{tx}}=13$ dB and the EH requirement $E_{s}=0$ dB, respectively. Fig. 2(a) shows the convergence of achieved SEE with Algorithm 1. Fig. 2(b) and 2(c) illustrate the convergence of parameter $\Delta F$ in Algorithm 1 and parameter $\Delta \eta$ in Algorithm 2, respectively. These two parameters control the termination of the iterative processes in both algorithms. As seen in these numerical results, the maximum SEE and the convergence of both algorithms can be achieved with a limited number of iterations.
\begin{center}
	\begin{figure}[ht!]
		\includegraphics[width=\linewidth]{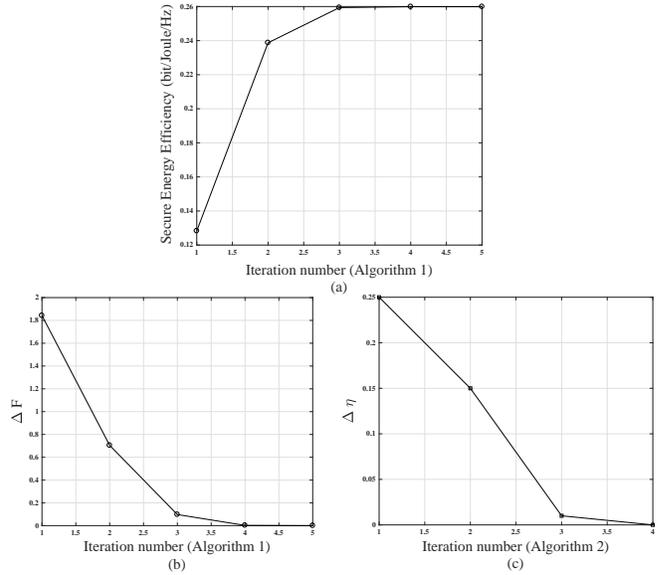}
		\caption{Convergence results of our proposed algorithms by these assumptions: $P_{\textrm{tx}}=13$ dB, $E_{s}=0$ dB and $R_{d} = 0.5$ bps/Hz}
	\end{figure}
\end{center}
\begin{center}
	\begin{figure}[ht!]
		\includegraphics[width=\linewidth]{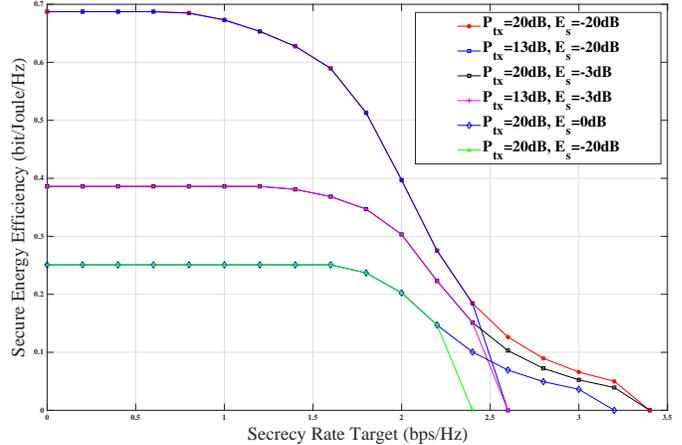}
		\caption{Achieved SEE with different target secrecy rates and transmit power constraints and harvest energy requirements}
	\end{figure}
\end{center}
Fig. 3 illustrates the achieved SEE with different target secrecy rates and EH requirements. As seen in Fig. 3, the optimal SEE decreases as the target rate increases. Note that the zero SEE means that problem is not feasible with a given target secrecy rate constraint. On the other hand, the SEE can achieve a better performance with a smaller EH requirement. In addition, if the problem is feasible with a given target secrecy rate constraint with small transmit power consumption, it would be able to achieve the same SEE with larger transmit power consumption. Increasing the transmit power consumption cannot yield a better SEE, however, it should be able to achieve a higher target secrecy rate.
\begin{center}
	\begin{figure}[ht!]
		\includegraphics[width=\linewidth]{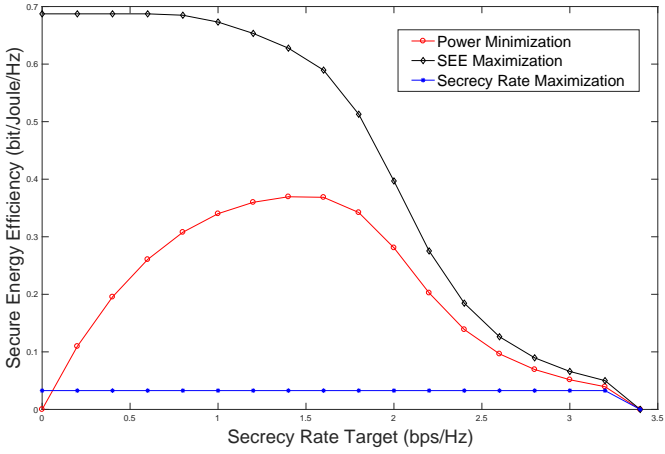}
		\caption{Achieved SEE for different schemes: SEE maximization, power minimization and secrecy rate maximization}
	\end{figure}
\end{center}
Fig. 4 compares the achievable SEE of three schemes: SEE maximization, power minimization and secrecy rate maximization. In these simulation results, the transmit power constraint is assumed to be 20 dB and the EH requirement is -20dB. As expected, the proposed scheme for SEE maximization achieves the best SEE of all the three schemes. As can be seen in this figure, the achievable SEE performance obtained from secrecy rate maximization is not affected by the target secrecy rate values in its feasible domain. This can be explained as follows. The power and energy limitations become major concerns in secrecy rate maximization problems, and therefore the limited power is used fully to maximize the secrecy rate. Hence, the ratio of secrecy rate and transmit power consumption does not change with a fixed transmit power constraint. Furthermore, the zero SEE means the target secrecy rate cannot be achieved with the available transmit power.
\section{Conclusion}
In this paper, we have considered the SEE maximization problem for an underlay MISO CR network. In particular, the transmit covariance matrix was designed to provide the required secrecy rate at the SU-Rx while satisfying the interference leakage constraint on the PU-Rx and the EH requirement on the ER. The original problem was not convex due to the non-linear fractional objective function. To overcome this non-convexity issue, we converted the original problem into a convex one by exploiting non-linear fractional and DC programming. Simulation results were provided to validate the convergence of the proposed algorithms and the performance of the proposed SEE based resource allocation technique. In addition, the achievable SEE in the developed scheme was compared with two alternative schemes.

\begin{appendix}
	\subsection*{Proof of Proposition \ref{proposition:rank_proof}}\label{proof_of_proposition}
First, we consider the Langrange function of problem (11):
\begin{align}
&\mathcal{L}(\mathbf{Q}_{s},\mathbf{Z},\alpha,\beta,\gamma,\mu)=-\{\log_{2}(1+\mathbf{h}_{s}^{H}\mathbf{Q}_{s}\mathbf{h}_{s})\nonumber\\
&-\log_{2}(1+\mathbf{h}_{e}^{H}\mathbf{Q}_{s}\mathbf{h}_{e})-\lambda[ \textrm{tr}(\mathbf{Q}_{s})+P_{c}]\}- \textrm{tr}(\mathbf{Z}\mathbf{Q}_{s})\nonumber\\ &-\alpha[\log_{2}(1+\mathbf{h}_{s}^{H}\mathbf{Q}_{s}\mathbf{h}_{s})-\log_{2}(1+\mathbf{h}_{e}^{H}\mathbf{Q}_{s}\mathbf{h}_{e})- R_{d}]\nonumber\\
&- \beta[\eta_{eh}[\mathbf{h}_{e}\mathbf{Q}_{s}\mathbf{h}_{e}+1] - E_{s}]+\gamma[\mathbf{h}_{p}^{H}\mathbf{Q}_{s}\mathbf{h}_{p}-P_{f}]\nonumber\\
&+\mu[\textrm{tr}(\mathbf{Q}_{s}) - P_{\textrm{tx}}]	
\end{align}
where $\mathbf{Q}_{s} \in \mathbb{H}^{N_{t}}_{+}$, $\mathbf{Z}\in \mathbb{H}^{N_{t}}_{+}$, $\alpha\in \mathbb{R}_{+}$, $\beta\in\mathbb{R}_{+}$, $\gamma\in\mathbb{R}_{+}$, $\mu\in\mathbb{R}_{+}$ are the Lagrangian multipliers associated with problem (11). Then we derive the corresponding  Karush-Kuhn-Tucker (KKT) conditions \cite{boyd2004convex}:
\begin{align}
&\frac{\partial\mathcal{L}}{\partial\mathbf{Q}_{s}}=-(\alpha+1)[\frac{\mathbf{h}_{s}^{H}\mathbf{h}_{s}}{(1+\mathbf{h}_{s}^{H}\mathbf{Q}_{s}\mathbf{h}_{s})\ln2}] + (1-\alpha-\beta\eta_{eh})\nonumber\\
&[
\frac{\mathbf{h}_{e}^{H}\mathbf{h}_{e}}{(1+\mathbf{h}_{e}^{H}\mathbf{Q}_{s}\mathbf{h}_{e})\ln2}]+(\lambda+\mu)\mathbf{I}-\mathbf{Z}-\gamma\frac{\mathbf{h}_{p}^{H}\mathbf{h}_{p}}{(1+\mathbf{h}_{p}^{H}\mathbf{Q}_{s}\mathbf{h}_{p})\ln2}\nonumber\\
&=0\\
&\mathbf{Z}\mathbf{Q}_{s}=0, \mathbf{Z}\succeq 0
\end{align}
The following equality holds:
\begin{align}
&-(\alpha+1)[\frac{\mathbf{h}_{s}^{H}\mathbf{h}_{s}}{(1+\mathbf{h}_{s}^{H}\mathbf{Q}_{s}\mathbf{h}_{s})\ln2}] + (1-\alpha-\beta\eta_{eh})\nonumber\\
&[\frac{\mathbf{h}_{e}^{H}\mathbf{h}_{e}}{(1\!+\!\mathbf{h}_{e}^{H}\mathbf{Q}_{s}\mathbf{h}_{e})\ln2}]\!+\!(\lambda\!+\!\mu)\mathbf{I}\!-\!\gamma\frac{\mathbf{h}_{p}^{H}\mathbf{h}_{p}}{(1\!+\!\mathbf{h}_{p}^{H}\mathbf{Q}_{s}\mathbf{h}_{p})\ln2}\!=\!\mathbf{Z}
\end{align}
\begin{align}
&\Rightarrow \{-(\alpha+1)[\frac{\mathbf{h}_{s}^{H}\mathbf{h}_{s}}{(1+\mathbf{h}_{s}^{H}\mathbf{Q}_{s}\mathbf{h}_{s})\ln2}] + (1-\alpha-\beta\eta_{eh})\nonumber\\
&[\frac{\mathbf{h}_{e}^{H}\mathbf{h}_{e}}{(1\!+\!\mathbf{h}_{e}^{H}\mathbf{Q}_{s}\mathbf{h}_{e})\ln2}]\!+\!(\lambda\!+\!\mu)\mathbf{I}\!-\!\gamma\frac{\mathbf{h}_{p}^{H}\mathbf{h}_{p}}{(1\!+\!\mathbf{h}_{p}^{H}\mathbf{Q}_{s}\mathbf{h}_{p})\ln2}\}\mathbf{Q}_{s}\nonumber\\
&=0
\end{align}
\begin{align}
&\Rightarrow 
\{ (1-\alpha-\beta\eta_{eh})[
\frac{\mathbf{h}_{e}^{H}\mathbf{h}_{e}}{(1+\mathbf{h}_{e}^{H}\mathbf{Q}_{s}\mathbf{h}_{e})\ln2}]+(\lambda+\mu)\mathbf{I}-\nonumber\\
&\gamma\frac{\mathbf{h}_{p}^{H}\mathbf{h}_{p}}{(1\!+\!\mathbf{h}_{p}^{H}\mathbf{Q}_{s}\mathbf{h}_{p})\ln2}\}\mathbf{Q}_{s}\!=\!\{(\alpha\!+\!1)[\frac{\mathbf{h}_{s}^{H}\mathbf{h}_{s}}{(1\!+\!\mathbf{h}_{s}^{H}\mathbf{Q}_{s}\mathbf{h}_{s})\ln2}]\}\mathbf{Q}_{s}
\end{align}
\begin{align}
&\Rightarrow 
\mathbf{Q}_{s}=\{(\alpha+1)[\frac{\mathbf{h}_{s}^{H}\mathbf{h}_{s}}{(1+\mathbf{h}_{s}^{H}\mathbf{Q}_{s}\mathbf{h}_{s})\ln2}]\}\{(1-\alpha-\beta\eta_{eh})\nonumber\\
&[\frac{\mathbf{h}_{e}^{H}\mathbf{h}_{e}}{(1\!+\!\mathbf{h}_{e}^{H}\mathbf{Q}_{s}\mathbf{h}_{e})\ln2}]\!+\!(\lambda\!+\!\mu)\mathbf{I}\!-\!\gamma\frac{\mathbf{h}_{p}^{H}\mathbf{h}_{p}}{(1\!+\!\mathbf{h}_{p}^{H}\mathbf{Q}_{s}\mathbf{h}_{p})\ln2}\}^{-1}\mathbf{Q}_{s}
\end{align}
Hence, the following rank relaltion holds:
\begin{align}
&\textrm{rank}(\mathbf{Q}_{s})\!=\!\textrm{rank}\{\{(\alpha\!+\!1)[\frac{\mathbf{h}_{s}^{H}\mathbf{h}_{s}}{(1\!+\!\mathbf{h}_{s}^{H}\mathbf{Q}_{s}\mathbf{h}_{s})\ln2}]\}\{(1\!-\!\alpha\!-\!\beta\eta_{eh})\nonumber\\
&[\frac{\mathbf{h}_{e}^{H}\mathbf{h}_{e}}{(1\!+\!\mathbf{h}_{e}^{H}\mathbf{Q}_{s}\mathbf{h}_{e})\ln2}]\!+\!
(\lambda\!+\!\mu)\mathbf{I}\!-\!\gamma\frac{\mathbf{h}_{p}^{H}\mathbf{h}_{p}}{(1\!+\!\mathbf{h}_{p}^{H}\mathbf{Q}_{s}\mathbf{h}_{p})\ln2}\}^{-1}\mathbf{Q}_{s}\}\nonumber\\
&\leq\textrm{rank}[\frac{\mathbf{h}_{s}^{H}\mathbf{h}_{s}}{(1+\mathbf{h}_{s}^{H}\mathbf{Q}_{s}\mathbf{h}_{s})\ln2}]\leq 1.
\end{align}
which completes the proof of proposition 1.	
\end{appendix}
\bibliographystyle{IEEEtran}

\bibliography{referenceIEEE}

\end{document}